\documentclass[prx,twocolumn,preprintnumbers,superscriptaddress,longbibliography]{revtex4-1}
\usepackage{amssymb}
\usepackage{amsmath}
\usepackage{amsfonts}
\usepackage{graphicx}
\usepackage{dcolumn}
\usepackage{bm}
\usepackage{verbatim}
\usepackage{color}
\usepackage{wasysym}
\usepackage{xcolor}
\usepackage{times}
\usepackage{soul}
\usepackage{braket}
\usepackage{natbib}
\usepackage{floatrow}
\usepackage{comment}
\usepackage{soul,xcolor}
\usepackage[normalem]{ulem}
\usepackage{ulem}
\usepackage{appendix}
\usepackage[colorlinks,
citecolor = blue,
linkcolor = blue,
urlcolor  = blue,
anchorcolor = blue,
breaklinks = true
]{hyperref}

\definecolor{black}{rgb}{0,0,0}
\definecolor{blue}{rgb}{0,0,1}
\definecolor{green}{rgb}{0,1,0}
\definecolor{red}{rgb}{1,0,0}
\definecolor{brown}{rgb}{0.4,0.2,0}
\definecolor{darkgreen}{rgb}{0,0.7,0}
\definecolor{darkblue}{rgb}{0.0,0.0,0.5}
\definecolor{red}{rgb}{1,0,0}
\definecolor{deepmagenta}{rgb}{0.8, 0.0, 0.8}

\begin{document}

\title{Quantum Phases for Finite-Temperature Gases of Bosonic Polar Molecules Shielded by Dual Microwaves}
\date{\today }
\author{Wei Zhang}
\affiliation{Institute of Theoretical Physics, Chinese Academy of Sciences, Beijing 100190, China} 
\affiliation{School of Physical Sciences, University of Chinese Academy of
Sciences, Beijing 100049, China}

\author{Kun Chen}
\email{chenkun@itp.ac.cn}
\affiliation{Institute of Theoretical Physics, Chinese Academy of Sciences, Beijing 100190, China} 
\affiliation{School of Physical Sciences, University of Chinese Academy of Sciences, Beijing 100049, China}

\author{Su Yi}
\email{yisu@nbu.edu.cn}
\affiliation{Institute of Fundamental Physics and Quantum Technology \& School of Physics Science and Technology, Ningbo University, Ningbo, 315211, China}
\affiliation{Peng Huanwu Collaborative Center for Research and Education, Beihang University, Beijing 100191, China}

\author{Tao Shi}
\email{tshi@itp.ac.cn}
\affiliation{Institute of Theoretical Physics, Chinese Academy of Sciences, Beijing 100190, China} 
\affiliation{School of Physical Sciences, University of Chinese Academy of
Sciences, Beijing 100049, China}

\begin{abstract}
We investigate the finite-temperature phase diagram of polar molecules shielded by dual microwave fields using the path integral Monte Carlo method combined with the worm algorithm. We determine the critical temperature $T_c$ for Bose-Einstein condensations (BECs) and identify two distinct phases below $T_c$: the expanding gas (EG) phase and the self-bound gas (SBG) phase. We further analyze the temperature and interaction-strength dependence of the condensate and superfluid fractions. Notably, in contrast to dilute atomic BECs, the SBG phase displays a low condensate fraction and a high superfluid fraction, resembling the behavior of strongly correlated $^4$He superfluids. These significant many-body correlations arise from the interplay between long-range dipole-dipole interactions and the short-range shielding potential. Furthermore, we demonstrate that the aspect ratio of the gas provides a characteristic geometric signature to accurately determine the EG-to-SBG transition, robust against external trapping potentials. Our findings provide unbiased and numerically exact results to guide upcoming experiments with polar molecules.
\end{abstract}

\maketitle

\textit{Introduction}.--- Recent advances in the field of ultracold polar molecules, enabled by innovative microwave-shielding techniques~\cite{Karman2018,Quemener2018,Ye2019,Luo2021,Doyle2021,Wang2023,Will2023}, have led to groundbreaking achievements. Notable examples include the realization of a degenerate Fermi gas of NaK molecules~\cite{Luo2022a,Luo2022b} and the creation of ultracold tetramers (NaK)$_2$~\cite{chen2023,Deng2024} using a single microwave field, as well as the achievement of Bose-Einstein condensations (BECs) in NaCs gases~\cite{Will2023b} via dual microwave fields. These milestones have firmly established ultracold polar molecules as a versatile platform for exploring novel quantum phenomena~\cite{You1999,Baranov2002,Baranov2004,Marenko2006,Shi2010,Hirsch2010,Pu2010,Shlyapnikov2011,Shi2014,Jin2024,Langen2024}, with a wide range of applications in quantum simulation~\cite{Zoller2006,Pfau2009,Baranov2012,Zwierlein2021}, quantum computing~\cite{DeMille2002,Cornish2020,Zoller2006a,Tesch2002,Wall2015,Albert2020}, ultracold chemistry~\cite{Krem2008,Ni2019,Liu2022}, and precision measurements~\cite{Kozlov2007,Berger2010,Hinds2011,ThO2014,ThO2018,Hutzler2020}.

The mechanisms underlying both single and dual microwave shielding techniques have been elucidated using effective potential models~\cite{Deng2023,Deng2025,Karman2025}, where microwave-shielded polar molecules (MSPMs) experience a negated dipole-dipole interaction (DDI) at long range and a $1/r^6$ shielding potential at short range. This extensive shielding core induces strong many-body correlations, setting MSPMs apart from traditional dipolar atoms. These correlations invalidate the conventional Gross-Pitaevskii equation and its extensions including the Lee-Huang-Yang corrections~\cite{Jin2024,Langen2024}, particularly in self-bound gas configurations. While the zero-temperature properties of MSPMs have been explored using cluster expansion methods~\cite{Jin2024,Deng2025} and path integral ground state Monte Carlo simulations~\cite{Langen2024}, the many-body phenomena at finite temperatures remain elusive. Given the ongoing experimental work with NaCs~\cite{Will2023b} and the anticipated studies on other bosonic MSPMs, such as NaRb~\cite{Wang2023}, a deeper understanding of the finite-temperature properties of MSPMs—beyond the realm of dipolar atomic BECs and $^4$He physics—is urgently needed.

In this Letter, we investigate the finite-temperature behavior of a trapped gas of bosonic MSPMs under realistic experimental conditions using the state-of-the-art Path Integral Monte Carlo (PIMC) method combined with the Worm Algorithm (WA)~\cite{prokof1998exact, Boninsegni2006a, Boninsegni2006b}. This highly efficient, unbiased, and numerically exact approach has been successfully applied to study a wide range of strongly correlated systems, including ultracold atomic gases~\cite{chen2014universal}, quantum magnets~\cite{chen2013deconfined}, and quantum gases with long-range interaction~\cite{he4, kuklov2024transverse, plasma2023}. In the finite-temperature phase diagram, we determine the critical temperature $T_c$ for BECs, which is enhanced (reduced) by the attractive (repulsive) interactions. Below $T_c$, we identify a phase transition between the expanding gas (EG) phase and the self-bound gas (SBG) phase, both of which exhibit significant many-body correlations characterized by anti-bunched density-density correlations~\cite{Jin2024,Deng2025}. We analyze two key quantities: the condensate fraction, governed by the interplay between long-range DDI and the short-range shielding potential, and the superfluid fraction, determined by low-lying phonon excitations. While both quantities increase as temperature decreases, their responses to changes in interaction strength differ fundamentally, in contrast to dilute atomic BECs. Notably, the SBG phase exhibits a low condensate fraction and a high superfluid fraction, resembling the behavior of strongly correlated $^4$He superfluids~\cite{Onsager,Sears1982,Sears1983,Sokol1990,Sokol1991}. Furthermore, we demonstrate that although radial confinement blurs the phase boundary measured by the gas volume, the aspect ratio of the gas provides a clear signature for accurately determining the EG-to-SBG transition, highlighting the intricate interplay between condensate fractions and DDI. 

\textit{Model}.--- We consider a trapped gas of $N$ bosonic polar molecules under dual microwaves at temperature $T$. In second-quantized form, the total Hamiltonian of the system reads
\begin{align}
H&=\int d{\bm{r}}\left[ \frac{\hbar^2}{2M}\nabla \hat{\psi}^{\dagger }({
\bm{r}})\nabla \hat{\psi}({\bm{r}})+U({\bm{r}})\hat{\psi}
^{\dagger }({\bm{r}})\hat{\psi}({\bm{r}})\right]\nonumber\\
&\quad+\frac{1}{2}\int d{\bm{r}}d{\bm{r}}^{\prime }V(
\bm{r}-\bm{r}^{\prime })\hat{\psi}^{\dagger }({\bm{r}})\hat{\psi}%
^{\dagger }({\bm{r}}^{\prime })\hat{\psi}({\bm{r}}^{\prime })\hat{%
\psi}({\bm{r}}),
\end{align}%
where $M$ is the molecular mass, $\hat{\psi}({\bm{r}})$ is the field operator, and $U({\bm{r}})=M[\omega _{\perp }^{2}(x^{2}+y^{2})+\omega _{z}^{2}z^{2}]/2$ is the trapping potential with $\omega _{\perp }$ and $\omega _{z}$ being the transverse and axial trap frequencies, respectively. The interaction potential between two MSPMs takes the form~\cite{Deng2025}
\begin{align}
V({\bm r})&=\frac{C_{6}}{r^{6}}\left[ \sin ^{4}\theta +\frac{w_{1}}{w_{2}}\sin ^{2}\theta \cos ^{2}\theta+\frac{w_{0}}{w_{2}}(3\cos ^{2}\theta-1)^{2}\right]\nonumber\\
&\quad+\frac{C_{3}}{r^{3}}\left( 3\cos ^{2}\theta
-1\right),  \label{Veff}
\end{align}
where $C_{6}$ ($>0$) is the strength of the shielding potential, $\theta$ is the polar angle of the relative coordinate ${\bm r}$, and $w_{m=0,1,2}$ quantify the contributions from the angular momentum partial waves $\left\vert Y_{2m}(\hat{\boldsymbol r})\right\vert ^{2}$, respectively. The first term on the right-hand side of Eq.~\eqref{Veff} is always positive, ensuring that the shielding potential is repulsive in all directions. Additionally, $C_3$ is the strength of the negated long-range DDI. We emphasize that all interaction parameters in Eq.~\eqref{Veff} are determined by the permanent dipole moment $d$ of the molecules, as well as by the Rabi frequencies ($\Omega_+, \Omega_\pi$) and detunings ($\delta_+, \delta_\pi$) of the $\sigma^+$- and $\pi$-polarized microwave fields. As demonstrated in Ref.~\cite{Deng2025}, the approximate analytic expressions for all interaction parameters yield scattering properties with high precision, validating their applicability to the study of many-body physics.

\begin{figure}[tbp]
\includegraphics[trim=50 0 80 5, clip, width=1\columnwidth]{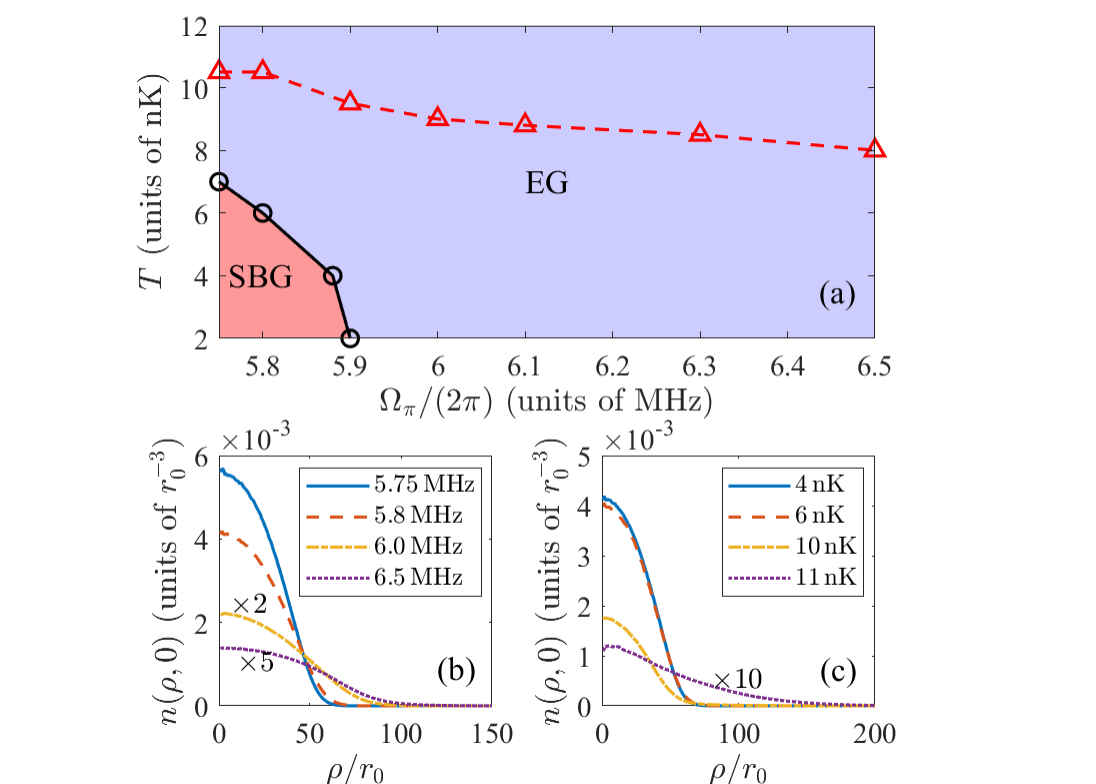}
\caption{(a) Phase diagram on the $\Omega_\pi$-$T$ plane. The dashed line denotes the critical temperature $T_c$ for the BEC transition and the solid line indicates the critical temperature $T_{\rm sb}$ for the SBG phase. (b) Radial density profiles at $T=4\,{\rm nK}$ for different $\Omega_\pi$'s. (c) Radial density profiles at $\Omega_\pi=\Omega_\pi^<$ for different $T$'s. To improve visibility, some density profiles in (b) and (c) are scaled by appropriate factors. All results presented here are for a system of $N=200$ molecules.}
\label{phase}
\end{figure}

As a concrete example, we consider NaCs molecules with a permanent dipole moment $d=4.6\,{\rm Debye}$. Following the experimental setup in Ref.~\cite{Will2023b}, we assume the Rabi frequency and detuning of the $\sigma^+$-polarized microwave field to be $\Omega _{+}=2\pi \times 7.9\,\mathrm{MHz}$ and $\delta _{+}=-2\pi \times 8\,\mathrm{MHz}$, respectively. Additionally, the detuning of the $\pi$-polarized microwave field is fixed at $\delta_{\pi}=-2\pi \times 10\,\mathrm{MHz}$, leaving $\Omega_\pi$ as the sole free parameter for tuning the inter-molecular potential. Throughout this work, we consider the range of the Rabi frequency $\Omega_\pi\in2\pi\times [5.75,6.5]\,{\rm MHz}$, within which $C_3$ remains non-negative. Consequently, DDI is always attractive in the $xy$ plane and repulsive along the $z$ axis. For convenience, we introduce two specific Rabi frequencies: $\Omega_\pi^>\equiv 2\pi\times 6.5\,{\rm MHz}$ and $\Omega_\pi^<\equiv 2\pi\times 5.8\,{\rm MHz}$. At $\Omega_\pi=\Omega_\pi^>$, $C_3\approx0$, and DDI is nearly canceled. As $\Omega_\pi$ decreases, $C_3$ monotonically increases, strengthening the long-range attraction in the $xy$ plane and reducing the size of the short-range shielding core. At $\Omega_\pi=\Omega_\pi^<$, the dipolar interaction becomes sufficiently strong. Thus, $\Omega_\pi^>$ and $\Omega_\pi^<$ are the representative Rabi frequencies for weak and strong dipolar interactions, respectively. We note that, as shown in Ref.~\cite{Deng2025}, no two-body bound states exist within the range of Rabi frequencies considered in this work. Finally, following the experimental conditions, the axial trap frequency is set to $\omega_z=2\pi\times 59\,{\rm Hz}$. Unless otherwise specified, the radial frequency is chosen as $\omega_\perp=\omega_\perp^>$ ($\equiv 2\pi\times33.6\,{\rm Hz}$), such that it matches the geometric average of the radial trap frequencies in the experiment~\cite{Will2023b}.

To investigate the finite-temperature many-body physics of this strongly correlated molecular gas, we employ the PIMC based on WA~\cite{prokof1998exact, Boninsegni2006a, Boninsegni2006b}. A key advantage of WA over conventional PIMC methods is its ability to sample an extended ensemble that includes both worldline configurations with fixed particle numbers (diagonal sector) and those with particle-number fluctuations (off-diagonal sector). This is accomplished through the introduction of ``worm" operators, which locally create and annihilate particles, thereby enabling the exploration of configurations with varying particle numbers and non-zero winding numbers in imaginary time. These off-diagonal updates are essential for accurately capturing emergent phenomena such as BECs and superfluidity, where off-diagonal long-range order and global phase coherence dictate the system's behavior. Consequently, WA provides an unbiased and numerically exact framework for investigating the thermodynamic properties of the system under realistic experimental conditions, circumventing the need for mean-field approximations or uncontrolled expansions.

\textit{Phase diagram}.--- Figure~\ref{phase}(a) displays the phase diagram on the $\Omega_\pi$-$T$ parameter plane for $N=200$. The dashed line denotes the critical temperature $T_c$ for BECs, determined using the criterion $f_c(T_c)\approx 5\%$. Here, $f_c(T)\equiv N_0(T)/N$ is the condensate fraction, where $N_0$ is the largest eigenvalue of the first-order correlation function $G_1(\bm{r},\bm{r}^{\prime}) \equiv\langle\hat{\psi}^{\dag}(\bm{r})\hat{\psi}(\bm{r}^\prime)\rangle$~\cite{SM}. While the choice of the threshold condensate fraction for BECs is somewhat arbitrary, slight variations in this threshold only lead to quantitative changes in $T_c$. Notably, $T_c$ monotonically increases as $\Omega_\pi$ decreases, a trend that can be understood by analyzing the inter-molecular potential, as discussed below. The solid line in Fig.~\ref{phase}(a) marks the boundary between the self-bound gas (SBG) and expanding gas (EG) phases. Since the two-body interaction considered here does not support two-body bound states~\cite{Deng2025}, the SBG phase arises purely from many-body effects. The solid line can also be interpreted as a critical temperature $T_{\rm sb}$, such that for a given $N$ and $\Omega_\pi$, the SBG phase forms only when $T<T_{\rm sb}$. Although the SBG may persist above $T_c$, we always find that $T_{\rm sb}<T_c$ within the parameter regime explored in our calculations. This observation may arise because the interaction strengths considered here are insufficient to stabilize the SBG phase at higher temperatures.

In Fig.~\ref{phase}(b), we compare radial density profiles for different values of $\Omega_\pi$ at $T=4\,{\rm nK}$, where $r_0=10^3a_{\mathrm{B}}$ with $a_{\mathrm{B}}$ being the Bohr radius. Here, the radial sizes of the gas in the SBG phase ($\Omega_\pi\leq \Omega_\pi^<$) are significantly smaller than those in the EG phase. However, this distinction may not hold in the presence of radial confinement. For instance, as illustrated in Fig.~\ref{phase}(c), for a fixed $\Omega_\pi=\Omega_\pi^<$, the gas at $T=10\,{\rm nK}$ abruptly shrinks just below the BEC transition temperature at $10.5\,{\rm nK}$, rather than at the EG-to-SBG transition around $6\,{\rm nK}$. Particularly, the radial size of the gas at $T=10\,{\rm nK}$ is very close to (of the same order of magnitude as) that of the SBG at $6\,{\rm nK}$, indicating that the external potential blurs the boundary of the EG-to-SBG transition. Therefore, to definitively identify a self-bound gas in the calculations, it is necessary to completely switch off the radial trap.

\begin{figure}[tbp]
\includegraphics[width=0.9\columnwidth]{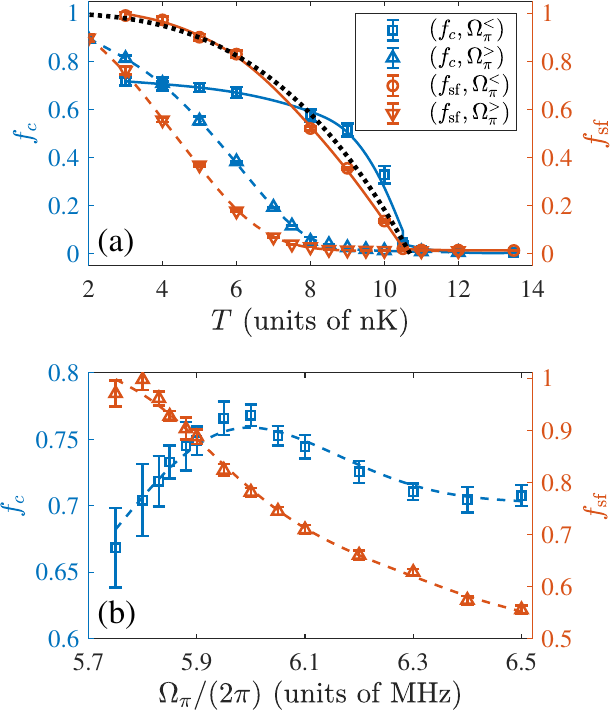}
\caption{(a) Condensate fraction $f_c$ (blue, left axis) and superfluid fraction $f_{\rm sf}$ (red, right axis) as functions of $T$ for $\Omega_\pi=\Omega_\pi^<$ ({solid} lines) and $\Omega_\pi^>$ (dash lines). The dotted line denotes the condensate fraction as a function of $T$ for a non-interacting gas. (b) Condensate fraction (left axis) and superfluid fraction (right axis) as functions of $\Omega_\pi$ at $T=4\,{\rm nK}$. For both (a) and (b), the system consists of $N=200$ molecules.}
\label{fvst}
\end{figure}

\textit{Thermodynamic transition}.--- To gain deeper insight into the thermodynamic phase transition, we plot the temperature dependence of the condensate fraction (blue lines) for the representative Rabi frequencies $\Omega_\pi=\Omega_\pi^<$ and $\Omega_\pi^>$ in Fig.~\ref{fvst}(a). Remarkably, we observe that for $\Omega_\pi=\Omega_\pi^>$, $f_c\approx 0.7$ and $0.9$ at $T=4\,{\rm nK}$ and $2\,{\rm nK}$, respectively, in agreement with the experimental measurement~\cite{Will2023b} and the cluster expansion result~\cite{Deng2025}. For comparison, we also plot $f_c$ as a function of $T$ (dashed black line) for a non-interacting gas in the same harmonic potential. Notably, the critical temperature for $\Omega_\pi=\Omega_\pi^<$ ($\Omega_\pi^>$) is higher (lower) than that of the non-interacting gas. This behavior can be understood as follows: when the two-body interaction is attractive, it is energetically favorable for molecules to occupy the same quantum state, thereby facilitating condensation. Roughly speaking, the stronger the attraction, the higher the critical temperature. Conversely, when the inter-molecular potential is repulsive, the critical temperature decreases, explaining why the critical temperature for $\Omega_\pi = \Omega_\pi^>$ is lower than that of the non-interacting gas.

Another frequently used physical quantity closely relating to the BEC transition is the superfluid fraction $f_{\rm sf}\equiv n_{\rm sf}/n$, where $n_{\rm sf}$ and $n$ are the superfluid and total densities of a homogeneous gas, respectively. For a finite system, $f_{\rm sf}$ can be evaluated by examining its response to an imposed rotation. This calculation is particularly convenient in PIMC, {as the total superfluid response along the axis of rotation is proportional to the square of the total projected area of the imaginary-time paths~\cite{PhysRevLett.63.1601}}. Physically, $f_{\rm sf}$ is primarily determined by the behavior of low-lying phonon modes, e.g., the sound velocity.

The red lines in Fig.~\ref{fvst}(a) show the temperature dependence of the superfluid fraction for different values of $\Omega_\pi$. As can be seen, $f_{\rm sf}$ monotonically increases to unity as $T$ decreases. Notably, the onset of a nonzero $f_{\rm sf}$ coincides with that of $f_c$, indicating that, in three-dimensional systems, the formation of BECs implies superfluidity. Indeed, the appearance of BECs signifies a change in the low-energy dispersion from the quadratic dispersion of free particles to the linear dispersion of phonons, which results in a finite Landau critical velocity and, consequently, superfluidity. We note that in lower dimensions, e.g., in two-dimensional systems, superfluidity emerges below the Berezinskii–Kosterlitz–Thouless transition temperature~\cite{Berezinskii,KT}.

\begin{figure}[tbp]
\includegraphics[width=1\columnwidth]{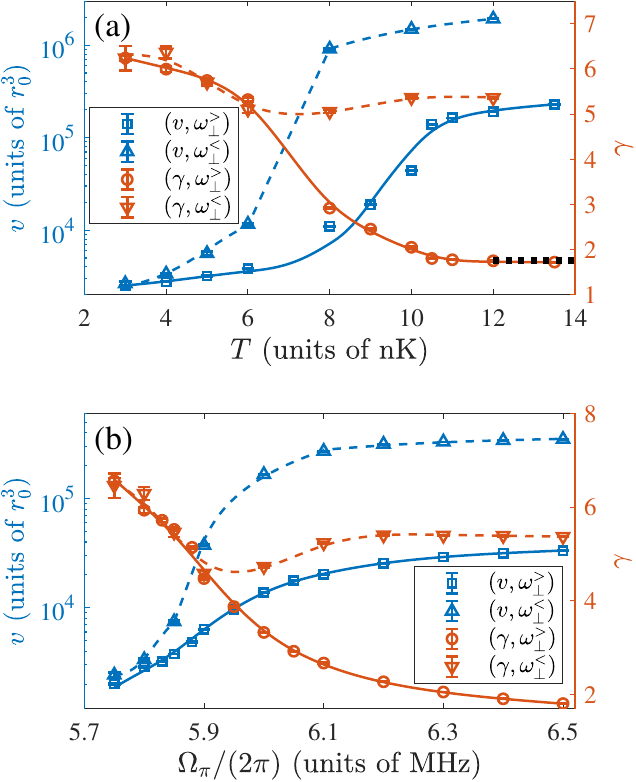}
\caption{(a) Gas volume $v$ (blue, left axis) and aspect ratio $\gamma$ (red, right axis) as functions of $T$ for $\omega_\perp=\omega_\perp^>$ ({solid} lines) and $\omega_\perp^<$ (dashed lines). The Rabi frequency is fixed at $\Omega_\pi=\Omega_\pi^<$. The dotted line denotes the aspect ratio of a trapped non-interacting classical gas. (b) Gas volume (left axis) and aspect ratio (right axis) as functions of $\Omega_\pi$ at $T=4\,{\rm nK}$. The solid (dashed) lines correspond to a higher (lower) radial trap frequency $\omega_\perp=\omega_\perp^>$ ($\omega_\perp^<$). For both (a) and (b), the number of molecules is $N=200$.}
\label{geovst}
\end{figure}

Furthermore, we explore the dependence of the condensate and superfluid fractions on the interaction strength ($\Omega_\pi$). Figure~\ref{fvst}(b) shows $f_c$ and $f_{\rm sf}$ as function of $\Omega_\pi$ for $T=4\,{\rm nK}$ and $N=200$. Unlike the temperature dependence observed in Fig.~\ref{fvst}(a), here $f_c$ and $f_{\rm sf}$ may exhibit contrasting behaviors with respect to $\Omega_\pi$. Specifically, while $f_{\rm sf}$ monotonically increases as $\Omega_\pi$ decreases, $f_c$ decreases in the small $\Omega_\pi$ regime. To understand this, we note that in the weak DDI regime (i.e., large $\Omega_\pi$), the gas has a lower density. Thus, decreasing $\Omega_\pi$ enhances the attraction, leading to a higher condensate fraction. However, in the strong dipolar interaction regime (i.e., small $\Omega_\pi$), the gas has a higher density, and the repulsive shielding potential becomes significant. As demonstrated in the Supplemental Material~\cite{SM}, this results in stronger anti-bunching in the density-density correlation functions. Therefore, further decreasing $\Omega_\pi$ increases depletion, reducing the condensate fraction. Thus, we conclude that in weak interaction (low density) regime, $f_c$ is mainly determined by the long-range (dipolar) interaction, whereas in the high-density regime, $f_c$ is dominated by the short-range shielding potential. The non-monotonic behavior of $f_c$ can be inferred from the momentum distribution~\cite{SM}, as measured in time-of-flight experiments~\cite{Will2023b}. In contrast, the superfluid fraction is closely related to the low-lying phonon excitations, which depend on the long-range behavior governed by the dipolar interaction. This explains why, as shown in Fig.~\ref{fvst}(b), $f_{\rm sf}$ is a monotonic function of $\Omega_\pi$.

\textit{EG-to-SBG transition}.--- Since the EG-to-SBG transition is not directly reflected in $f_c$ and $f_{\rm sf}$, we instead examine the geometric properties of the gases, such as its `volume' $v\equiv\sqrt{\langle x^2\rangle\langle y^2\rangle\langle z^2\rangle}$ and aspect ratio $\gamma\equiv\sqrt{\langle x^2\rangle/\langle z^2\rangle}$. Although the EG-to-SBG transition point is independent of the radial trap, the presence of radial confinement obscures the phase boundary, as the geometric properties of the gas strongly depend on $\omega_\perp$. To make the transition more apparent, we introduce a weaker radial trap with $\omega_\perp=\omega_\perp^<$ ($\equiv 2\pi\times 10.6\,{\rm Hz}$). In Fig.~\ref{geovst}(a), the blue lines show the volume as a function of temperature for two different values of $\omega_\perp$. As expected, the volume $v$ monotonically decreases as the temperature is lowered. Particularly, when $T$ becomes sufficiently small, the two volumes converge to approximately the same value, signaling the formation of SBG. In fact, as long as the radial size of SBG is smaller than that of the harmonic oscillator, the volumes of the gases corresponding to different radial trap frequencies should converge for control parameters deep within the SBG phase. However, it is important to noted that, due to the radial trap, the point where two volumes converge does not coincide with the EG-to-SBG transition.

Next, we consider the aspect ratio of the gases. To this end, we note that, for a non-interacting gas, $\gamma$ is $\omega_z/\omega_\perp$ for a classical gas at high temperatures and $\sqrt{\omega_z/\omega_\perp}$ for a pure condensate in the low-temperature limit. Therefore, for $\omega_z/\omega_\perp>1$, $\gamma$ decreases as the temperature is lowered. In the presence of DDI, however, the gas tends to stretch (measured in terms of aspect ratio) along the attractive (radial) directions to minimize the interaction energy. As a result, $\gamma$ increases with the interaction strength for a negated DDI. In summary, $\gamma$ is determined by the competition between condensation and interaction: condensation (interaction) favors a smaller (larger) $\gamma$.

The red lines in Fig.~\ref{geovst}(a) show the temperature dependence of $\gamma$ for two values of $\omega_\perp$ with $\Omega_\pi=\Omega_\pi^<$. In the high temperature limit, the aspect ratios agree with those of non-interacting classical gases. As temperature is lowered, $\gamma$ for the trap with $\omega_\perp=\omega_\perp^>$ monotonically increases, indicating that DDI dominates the behavior of $\gamma$. This result is expected because the higher gas density in a stronger radial trap enhances the role of interactions. For $\omega_\perp=\omega_\perp^<$, the temperature dependence of $\gamma$ becomes more complex and non-monotonic. Starting from the high-temperature regime, $\gamma$ initially decreases as $T$ is lowered, suggesting that condensation dominates the behavior of $\gamma$ in the low-density regime. As $T$ is further reduced, the gas density increases, enhancing the role of DDI, and $\gamma$ begins to increase with decreasing $T$. Moreover, $\gamma$ for different values of $\omega_\perp$ converges at sufficiently small $T$, similar to the behavior of $v$. However, unlike $v$, the temperature at which $\gamma$ converges is approximately $T_{\rm sb}$, indicating that $\gamma$ can be used to accurately determine the EG-to-SBG transition even in the presence of radial confinement.

To gain deeper insight into the geometric properties of the gas, we plot in Fig.~\ref{geovst}(b) the dependence of $v$ and $\gamma$ on $\Omega_\pi$ for two radial trap frequencies, $\omega_\perp=\omega_\perp^>$ and $\omega_\perp^<$. Since a smaller Rabi frequency enhances the attractive interaction, $v$ monotonically decreases as $\Omega_\pi$ is reduced. Similarly, two curves for the volume converges to approximately the same value in the SBG phase. Regarding the aspect ratio, we observe distinct behaviors for the two trap frequencies. For $\omega_\perp=\omega_\perp^>$, $\gamma$ monotonically increases as $\Omega_\pi$ is lowered, indicating that the gas density is sufficiently high, and the aspect ratio is primarily governed by the interaction. In contrast, for $\omega_\perp=\omega_\perp^<$, $\gamma$ exhibits non-monotonic behavior. With a weaker radial trapping potential, the gas density remains low, and the aspect ratio is dominated by condensation effects in the large \( \Omega_\pi \) regime. Consequently, $\gamma$ first decreases as $\Omega_\pi$ is reduced. However, when $\Omega_\pi$ falls below approximately $2\pi\times 5.9\,{\rm MHz}$, the gas density is sufficiently large in the SBG phase, such that the interaction becomes the decisive factor determining the aspect ratio. As a result, $\gamma$ begins to increase as $\Omega_\pi$ is further lowered. Again, the transition point identified through the convergence of $\gamma$ is more accurate than that via determined via the convergence of $v$. This highlights the sensitivity of the aspect ratio to the underlying strong-correlation mechanism governing the EG-to-SBG transition.


\textit{Discussions and Conclusions}.--- In conclusion, we employ the PIMC method combined with WA to study the finite temperature phase diagram of bosonic MSPMs, identifying both the EG and SBG phases. These phases exhibit many-body correlation effects beyond those observed in atomic BECs, e.g., intriguing behaviors in the condensate and superfluid fractions, as well as anti-bunched density-density correlations, which arise from the competition between the long-range DDI and the short-range shielding potential. Although radial confinement blurs the phase boundary measured by the volume, we demonstrate that the aspect radio of the density profile serves as a robust quantity to accurately determine the EG-to-SBG transition, reflecting the interplay between condensate fractions and DDI. As an outlook, the attractive DDI can be enhanced either in the $xy$-plane or along the $z$-direction by tuning the Rabi frequency and the ellipticity angle of the microwave field~\cite{Luo2022b,chen2023}, potentially leading to the emergence of supersolid phases~\cite{supersolid2019}. Our findings highlight the strong correlation effects in MSPMs at finite temperatures and provide unbiased, numerically exact simulations to guide upcoming experiments with polar molecules such as NaRb and CaF.

\textit{Acknowledgments}.--- This work was supported by the NSFC (Grants No. 12135018 and No.12047503), by National Key Research and Development Program of China (Grant No. 2021YFA0718304), and by CAS Project for Young Scientists in Basic Research (Grant No. YSBR-057).

\bibliography{ref_dMolecule}

\clearpage

\widetext

\begin{center}
\textbf{\large Supplemental Materials: Quantum Phases of Finite-Temperature
Bosonic Polar Molecules Shielded by Dual Microwaves}
\end{center}

This Supplemental Material is organized as follows. In Sec.~\ref{EOM} we provide
a detailed derivation of the first order correlation function for
non-uniform systems, which is used to determine the condensate fraction and momentum
distribution. In Sec.~\ref{g2}, we present the density-density correlation function.

\setcounter{equation}{0} \setcounter{figure}{0} \setcounter{table}{0} %
\setcounter{page}{1} \setcounter{section}{0} \makeatletter
\renewcommand{\theequation}{S\arabic{equation}} \renewcommand{\thefigure}{S%
\arabic{figure}} \renewcommand{\bibnumfmt}[1]{[S#1]} \renewcommand{%
\citenumfont}[1]{S#1} \renewcommand{\thesection}{S\arabic{section}}%
\setcounter{secnumdepth}{3}

\section{First order correlation functions in non-uniform systems}

\label{EOM}

In this section, we present a detailed computation of the first order
correlation function, $G_{1}(\bm{r},\bm{r}^{\prime })=\langle \hat{\psi}%
^{\dag }(\bm{r})\hat{\psi}(\bm{r}^{\prime })\rangle $, in a non-uniform
system using PIMC method combined with WA. The condensate fraction is determined by the largest eigenvalue of $G_{1}(\bm{r},\bm{r}^{\prime})$, and the
momentum distribution is obtained from its Fourier transform.

The first order correlation function, $G_{1}(\bm{r},\bm{r}^{\prime })=\langle 
\hat{\psi}^{\dag }(\bm{r})\hat{\psi}(\bm{r}^{\prime })\rangle $, satisfies
the following properties: (1) Normalization condition: $\mathrm{Tr}%
G_{1}\equiv \int d\bm{r}G_{1}(\bm{r},\bm{r})=N$, where $N$ denotes the total
particle number. (2) Positivity: The eigenvalues of $G_{1}$ are
non-negative. (3) Hermiticity: $G_{1}(\bm{r},\bm{r}^{\prime })=G_{1}^{\ast }(%
\bm{r}^{\prime },\bm{r})$. The eigenvalues $\lambda _{n}$ of $G_{1}$ are determined by the eigen-equation%
\begin{equation}
\int d\bm{r}^{\prime }G_{1}(\bm{r},\bm{r}^{\prime })\phi _{n}(\bm{r}^{\prime
})=\lambda _{n}\phi _{n}(\bm{r}),
\end{equation}%
where the eigenvalues are sorted in descending order, $\lambda _{0}\geq
\lambda _{1}\geq \lambda _{2}\geq ...\geq 0$, and satisfy the normalization
condition $\sum_{i}\lambda _{i}=N$. The condensate fraction is then defined
as $f_{\mathrm{c}}=N_{0}/N$, where $N_{0}\equiv \lambda _{0}$.

To efficiently compute $G_{1}$ through sampling, we first select an appropriate
set of basis functions:%
\begin{eqnarray}
g_{n_{\rho },m,n_{z}}^{(\mathrm{e})}(\rho ,\varphi ,z) &=&\frac{1}{\sqrt{%
\mathcal{N}_{n_{\rho }mn_{z}}^{(1)}}}\cos (m\varphi )\psi _{n_{\rho
},m}(\rho /\xi _{\rho })\phi _{n_{z}}(z/\xi _{z}),\ m=0,1,2,\cdots,   \notag
\\
g_{n_{\rho },m,n_{z}}^{(\mathrm{o})}(\rho ,\varphi ,z) &=&\frac{1}{\sqrt{%
\mathcal{N}_{n_{\rho }mn_{z}}^{(2)}}}\sin (m\varphi )\psi _{n_{\rho
},m}(\rho /\xi _{\rho })\phi _{n_{z}}(z/\xi _{z}),\ m=1,2,3,\cdots. 
\end{eqnarray}%
Here, $m$ is the quantum number of the projected angular
momentum along the $z$-direction, and $\bm r=(\rho ,z,\varphi )$ with $\rho $, $\varphi$,
and $z$ representing the radial coordinate, azimuthal angle, and axial coordinate, respectively. The functions $\psi
_{n_{\rho },m}(x)=x^{|m|}e^{-x^{2}/2}L_{n_{\rho }}^{(|m|)}(x^{2})$ and $\phi
_{n_{z}}(x)=e^{-x^{2}/2}H_{n_{z}}(x)$ are
eigenfunctions of the two- and one-dimensional harmonic oscillators, respectively,expressed in terms of the generalized Laguerre
polynomials $L_{n}^{(\alpha )}(x)$ and Hermite polynomials $H_{n}(x)$.
The characteristic lengths $\xi _{\rho }$ and $\xi _{z}$
should be chosen to match the spatial widths of the density profile in the $xy$ plane
and along the $z$-axis, ensuring rapid convergence as the cut-offs for $n_{\rho}$ and $n_{z}$ are increased. The corresponding
normalization factors are%
\begin{equation}
\mathcal{N}_{n_{\rho }mn_{z}}^{(1)}=s_{m}\pi ^{3/2}\frac{\xi _{\rho
}^{2}(n_{\rho }+|m|)!}{2n_{\rho }!}2^{n_{z}}n_{z}!\xi _{z}, \\
\mathcal{N}_{n_{\rho }mn_{z}}^{(2)}=\pi ^{3/2}\frac{\xi _{\rho }^{2}(n_{\rho
}+|m|)!}{2n_{\rho }!}2^{n_{z}}n_{z}!\xi _{z},
\end{equation}%
where $s_{m}=2$ for $m=0$ and $s_{m}=1$ for $m>0$.

By projecting $G_{1}(\bm{r},\bm{r}^{\prime })$ onto the space spanned by the orthonormal basis $g_{n_{\rho },m,n_{z}}^{(s=\mathrm{e,o})}(\rho ,\varphi ,z)$, we
obtain%
\begin{eqnarray}
(G_{1})_{(n_{\rho }mn_{z}),(n_{\rho }^{\prime }m^{\prime }n_{z}^{\prime
})}^{(s,s^{\prime })} &=&\int d\bm{r}\int d\bm{r}^{\prime }g_{n_{\rho
},m,n_{z}}^{(s)}(\rho ,\varphi ,z)G_{1}(\bm{r},\bm{r}^{\prime })g_{n_{\rho
}^{\prime },m^{\prime },n_{z}^{\prime }}^{(s^{\prime })}(\rho ^{\prime
},\varphi ^{\prime },z^{\prime })  \notag \\
&=&\frac{\int d\bm{r}\int d\bm{r}^{\prime }g_{n_{\rho },m,n_{z}}^{(s)}(\rho
,\varphi ,z)\mathrm{Tr}[e^{-\beta \hat{H}}\hat{\psi} ^{\dag }(\bm{r})\hat{\psi} (\bm{r}%
^{\prime })]g_{n_{\rho }^{\prime },m^{\prime },n_{z}^{\prime }}^{(s^{\prime
})}(\rho ^{\prime },\varphi ^{\prime },z^{\prime })}{\mathrm{Tr}[e^{-\beta 
\hat{H}}]}  \notag \\
&=&\frac{\int d\bm{r}\int d\bm{r}^{\prime }\mathrm{Tr}[e^{-\beta \hat{H}%
}\hat{\psi} ^{\dag }(\bm{r})\hat{\psi} (\bm{r}^{\prime })]g_{n_{\rho
},m,n_{z}}^{(s)}(\rho ,\varphi ,z)g_{n_{\rho }^{\prime },m^{\prime
},n_{z}^{\prime }}^{(s^{\prime })}(\rho ^{\prime },\varphi ^{\prime
},z^{\prime })}{\int d\bm{r}\int d\bm{r}^{\prime }\mathrm{Tr}[e^{-\beta \hat{%
H}}\hat{\psi} ^{\dag }(\bm{r})\hat{\psi} (\bm{r}^{\prime })]}  \notag \\
&&\times \frac{\int d\bm{r}\int d\bm{r}^{\prime }\mathrm{Tr}[e^{-\beta \hat{H%
}}\hat{\psi} ^{\dag }(\bm{r})\hat{\psi} (\bm{r}^{\prime })]}{\mathrm{Tr}[e^{-\beta \hat{H%
}}]}.  \label{G1}
\end{eqnarray}%
In this new basis, $G_{1}(\bm{r},\bm{r}^{\prime })$ can be efficiently
computed using WA. The first factor in the last equality of Eq.~\eqref{G1} can be sampled in the off-diagonal sector (when the worm is present),
while the second factor represents the relative probability between the
diagonal and off-diagonal sectors.

For a system with cylindrical symmetry along the $z$-axis, $G_{1}$ is
block diagonal, i.e., $m=m^{\prime}$. By diagonalizing $G_{1}$ in different
sectors $m$, we obtain the largest eigenvalue $\lambda _{0}$ and the
condensate fraction $f_{c}$. The convergence of all results is verified by systematically increasing the cut-offs for $n_{\rho}$ and $n_{z}$.

\begin{figure}[tbp]
\includegraphics[width=0.9\columnwidth]{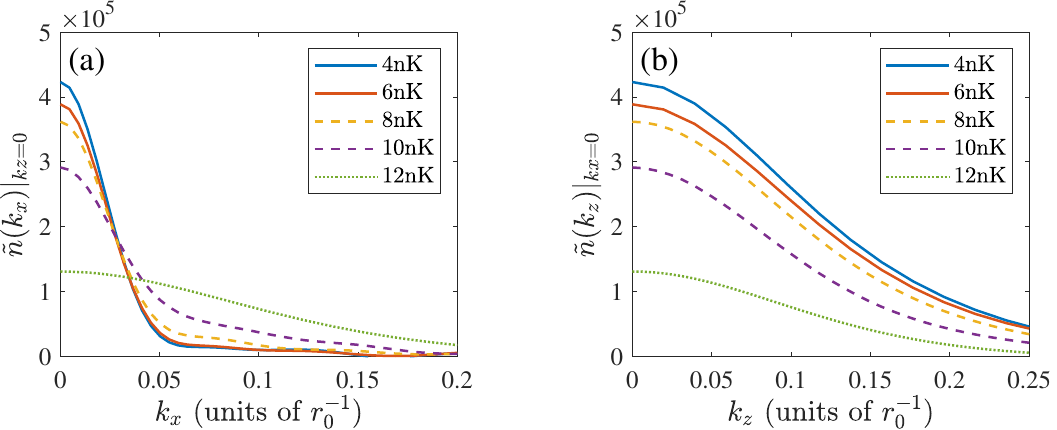}
\caption{\textbf{Integrated momentum distributions} The left panel
shows $\tilde{n}(k_x)|_{k_z=0} \equiv \tilde{n}(k_x, k_z=0)$, while the
right panel displays $\tilde{n}(k_z)|_{k_x=0} \equiv \tilde{n}(k_x=0, k_z)$.
The fixed parameters are $\Omega_{\protect\pi} = 2\protect\pi \times 5.8$
MHz, $N = 200$, and trap frequencies $\protect\omega_{\perp} = 2\protect\pi %
\times 33.6$ Hz, $\protect\omega_{z} = 2\protect\pi \times 59$ Hz. For this
parameter set, $T = 4$ nK and $T = 6$ nK correspond to the SBG phase, while
higher temperatures correspond to the EG phase. The superfluid transition
occurs at approximately $T = 10.5$ nK.}
\label{nkxkz}
\end{figure}

The momentum distribution is obtained from the Fourier transform of $G_{1}(%
\bm{r},\bm{r}^{\prime })$ as%
\begin{equation}
\tilde{n}(\bm{k})=\int d\bm{r}\int d\bm{r}^{\prime }G_{1}(\bm{r},\bm{r}%
^{\prime })e^{-i\bm{k}\cdot (\bm{r}-\bm{r}^{\prime })}.
\end{equation}%
The integrated momentum distribution $\int dk_{y}\Tilde{n}(\bm{k})/(2\pi)$ is shown
in Fig. \ref{nkxkz}.

\section{Density-density correlation functions}
\label{g2}

\begin{figure}[tbp]
\includegraphics[width=0.9\columnwidth]{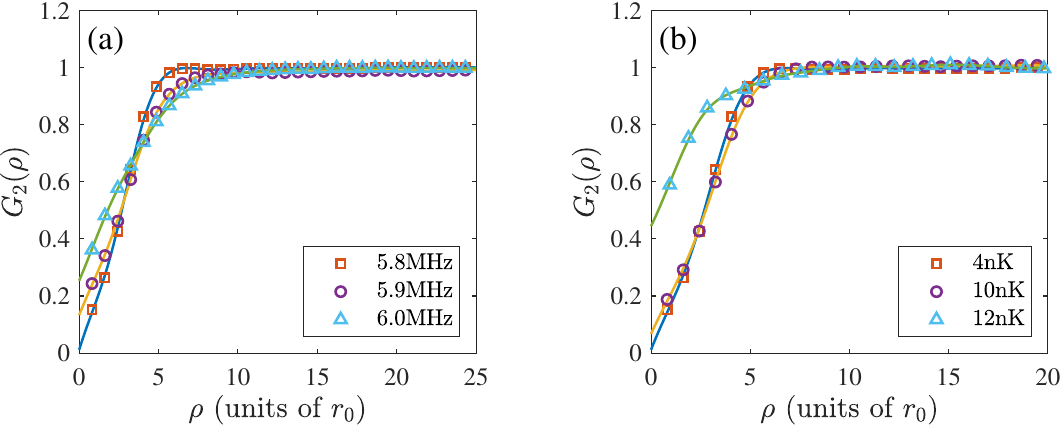}
\caption{\textbf{Density-density correlation functions.} The left panel
corresponds to $T = 4$ nK, while the right panel corresponds to $\Omega_{%
\protect\pi} = 2\protect\pi \times 5.8$ MHz, The fixed parameters are $N =
200$ and trap frequencies $\protect\omega_{\perp} = 2\protect\pi \times 33.6$
Hz, $\protect\omega_{z} = 2\protect\pi \times 59$ Hz. 
}
\label{G2}
\end{figure}

The density-density correlation function in the $xy$ plane is defined as%
\begin{equation}
G_{2}(\rho )\equiv \frac{\langle \sum_{i\neq j}\int dz\int dz^{\prime
}\delta (\bm{r}_{i}-\bm{r})\delta (\bm{r}_{j})\rangle }{\bar{n}(\rho )\bar{n}%
(0)},
\end{equation}%
where $\bm{r}_{i}$ is the position of the particle $i$ and $\bar{n}(\rho )=\int dzn(\bm{r})$ is the integrated density
distribution. As the dipolar interaction strength increases or the temperature
decreases, the mean interparticle distance is reduced, causing the short range
shielding potential to become more significant. Consequently, $G_{2}(\rho )$
exhibits stronger anti-bunching behavior for more attractive DDI and lower
temperatures, as shown in Fig. \ref{G2}.

\end{document}